\documentclass[12pt]{article}
\usepackage{epsfig}

\def\mean#1{\left<#1\right>}
\def\Journal#1#2#3#4{{#1}{\bf #2}, #3 (#4)}

\def\JPG{{J. Phys}~{G}}

\def\NPA{{Nucl. Phys. A}}
\def\NPB{{Nucl. Phys. B}}
\def\PLB{{Phys. Lett. B}}

\def\PRL{Phys. Rev. Lett.\ }

\def\PRD{{Phys. Rev. D}}
\def\PRC{{Phys. Rev. C}}

\def\ZPC{{Z. Phys. C}}
\def\ARNPS{{Ann. Rev. Nucl. Part. Sci.\ }} 

\def\Title#1{\begin{center} {\Large {\bf #1} } \end{center}}

\begin{document}

\Title{Hard Scattering and QCD Fundamentals at RHIC}

\bigskip\bigskip


\begin{raggedright}  

{\it Michael J. Tannenbaum\footnote{Research supported by U.S. Department of Energy, DE-AC02-98CH10886.}\index{Tannenbaum, M. J.}\\
Department of Physics\\
Brookhaven National Laboratory\\
Upton, NY 11973-5000, USA}
\bigskip\bigskip
\end{raggedright}
\section{Introduction}
  In 1998, at the QCD workshop in Paris, Rolf Baier asked me whether jets could be measured in Au+Au collisions because he had a prediction of a QCD medium-effect on color-charged partons traversing a hot-dense-medium composed of unscreened color-charges~\cite{BaierQCD98}. I told him~\cite{MJTQCD98} that there was a general consensus that for Au+Au central collisions at $\sqrt{s_{NN}}=200$ GeV, leading particles are the only way to measure jets, because in one unit of the nominal jet-finding cone,  $\Delta r=\sqrt{(\Delta\eta)^2 + (\Delta\phi)^2}$, there is an estimated $\pi\times{1\over {2\pi}} {dE_T\over{d\eta}}\sim 375$ GeV of energy !(!) The good news was that hard-scattering in p-p collisions was originally observed by the method of leading particles. 
  
  	The other good news was that the PHENIX detector had been designed to make such measurements and could identify and separate direct single $\gamma$ and $\pi^0$ out to $p_T\geq 30$ GeV/c~\cite{Shura2000}. It is ironic that the identification of $\pi^0$ and the separation from direct single $\gamma$ out to such a large $p_T$ in PHENIX was primarily driven by the desire to measure the polarized gluon structure function in p-p collisions in the range $0.10\leq x_T\leq 0.30$ via the longitudinal two-spin asymmetry of direct photon production~\cite{MJTQCD98}. This  illustrates that a good probe of QCD in a fundamental system such as p-p collisions, also provides a well calibrated probe of QCD in more complicated collisions such as Au+Au. In the 1980's when RHIC was proposed, hard processes were not expected to play a major role in A+A collisions. However, in 1998, inspired by Rolf and collaborators, and before them by the work of Gyulassy~\cite{MGyulassy} and Wang~\cite{XNWang}, I indicated~\cite{MJTQCD98} that my best bet on discovering the QGP was to utilize semi-inclusive $\pi^0$ or $\pi^{\pm}$ production in search for ``high $p_T$ suppression".  
\subsection{1998-How everything you want to know about jets can be done with leading particles. }
Following the discovery of hard-scattering in p-p collisions at the CERN-ISR~\cite{Darriulat} by the observation of an unexpectedly large yield of particles with large transverse momentum $(p_T)$, which proved that the quarks of DIS were strongly interacting, the attention of experimenters turned to measuring the predicted di-jet structure of the hard-scattering events using two-particle correlations. 
\begin{figure}[!htb]
\begin{center}
\begin{tabular}{cc}
\psfig{file=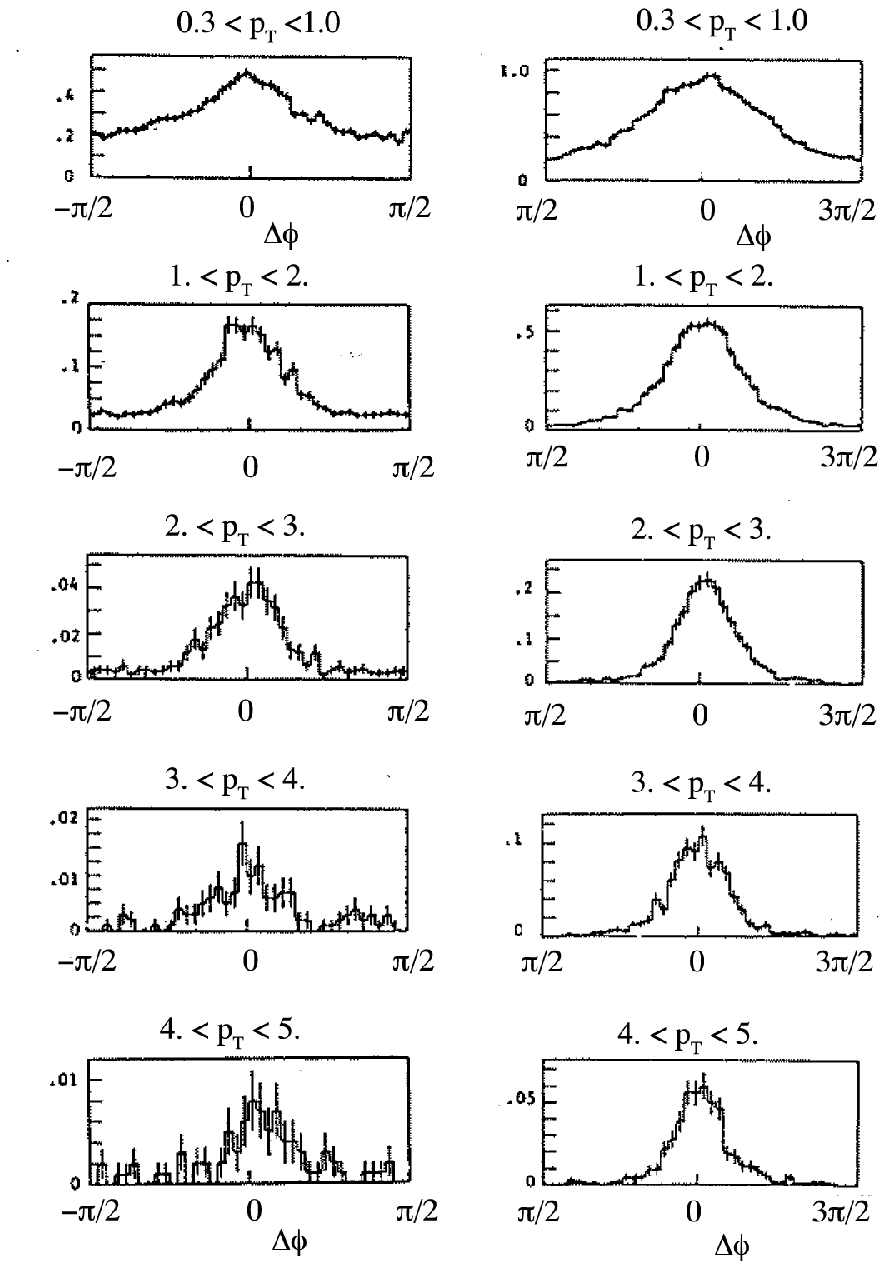,width=0.4\linewidth,height=0.5\linewidth}&
\psfig{file=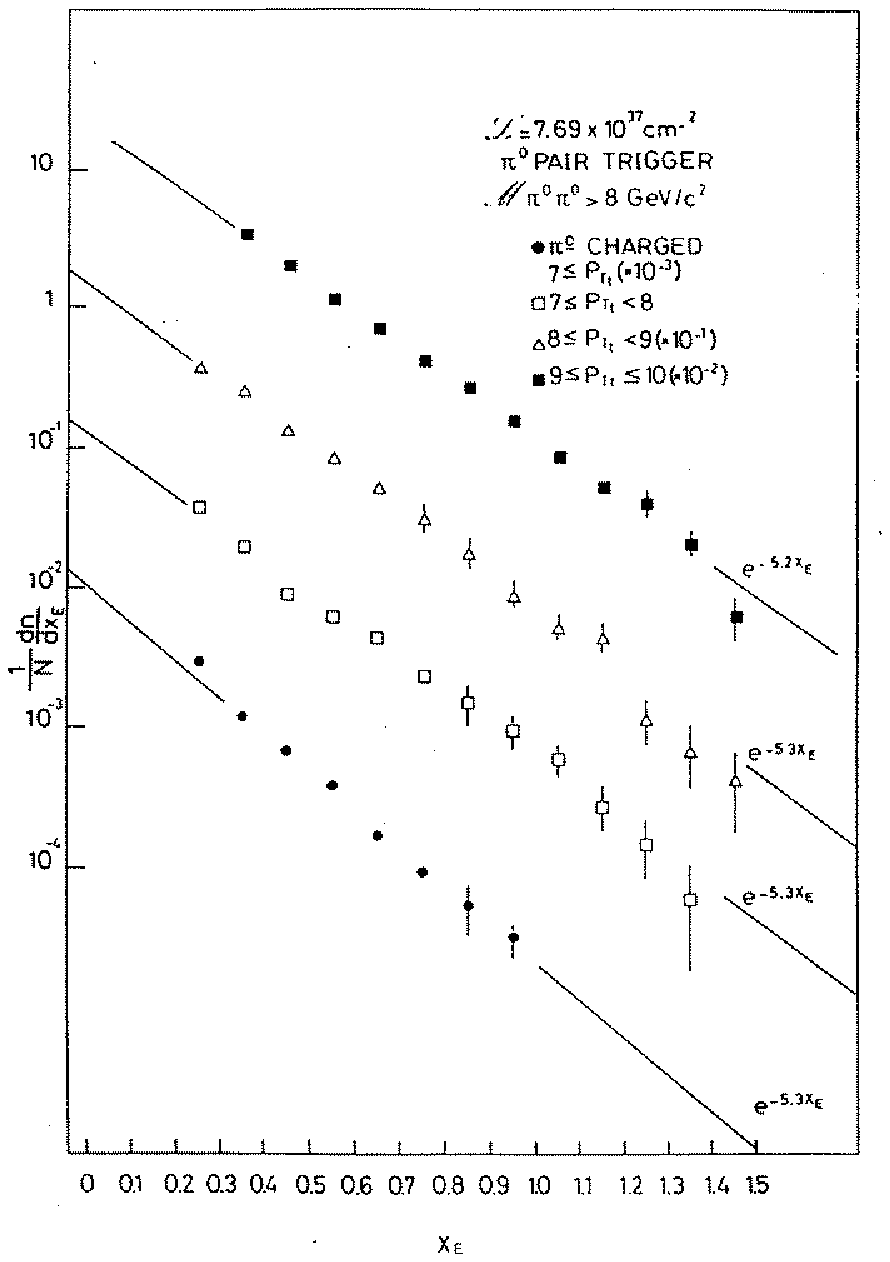,width=0.4\linewidth,height=0.5\linewidth}
\end{tabular}
\end{center}\vspace*{-0.15in}
\caption[]{a) (left) trigger-side b) (center) away-side correlations of charged particles with indicated $p_T$ for $\pi^0$ triggers with $p_{T_t} > 7$ GeV/c  . c) (right) $x_E$ distributions from this data .    \label{fig:mjt-ccorazi}}
\end{figure}
    The CCOR experiment~\cite{Angelis79}, using a $\pi^0$ trigger with transverse momentum $p_{T_t} > 7$ GeV/c,  was the first to provide associated charged particle measurement with full and uniform acceptance over the entire azimuth, with pseudorapidity coverage $-0.7\leq\eta\leq +0.7$, so that the jet structure of high $p_T$ scattering could be easily seen and measured (Fig.~\ref{fig:mjt-ccorazi}a,b). In all cases strong correlation peaks on flat backgrounds are clearly visible for both the trigger-side and the away-side, indicating di-jet structure.  The small variation of the widths of the away-side peaks for $p_{T}>1$ GeV/c (Fig.~\ref{fig:mjt-ccorazi}b) indicates out-of-plane activity of the di-jet system beyond simple jet fragmentation.

      	Following the methods of previous CERN-ISR experiments~\cite{Darriulat}  and the best theoretical guidance~\cite{FFF}, the away jet azimuthal angular distributions  of Fig.~\ref{fig:mjt-ccorazi}b, which were thought to be unbiased, were analyzed in terms of the two variables: $p_{\rm out}=p_T \sin(\Delta\phi)$, the out-of-plane transverse momentum of a track,   
 and $x_E$, where \\ 
\hspace*{0.05\linewidth}\begin{minipage}[b]{0.45\linewidth}
\[ 
x_E=\frac{-\vec{p}_T\cdot \vec{p}_{T_t}}{|p_{T_t}|^2}=\frac{-p_T \cos(\Delta\phi)}{p_{T_t}}\simeq \frac {z}{z_{t}}  
\]
\vspace*{0.001in}
\end{minipage}
\hspace*{0.02\linewidth}
\begin{minipage}[b]{0.39\linewidth} 
\vspace*{0.003in}
\psfig{file=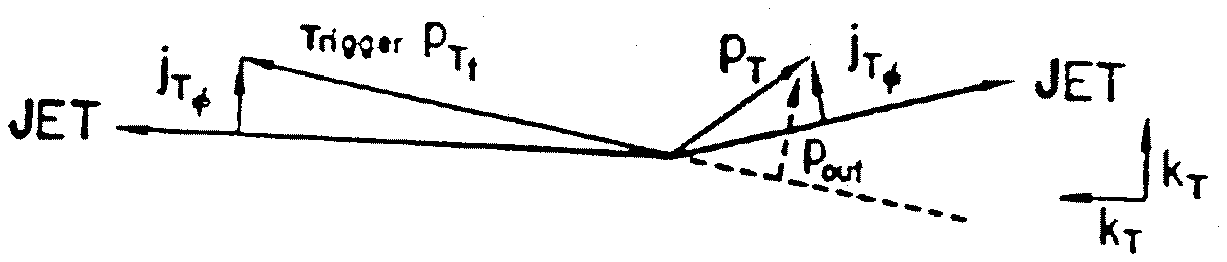, width=\linewidth}
\vspace*{-0.10in}
\label{fig:mjt-poutxe}
\end{minipage}
\vspace*{-0.12in}

\begin{figure}[!htb]
\begin{center}\vspace*{-0.12in}
\begin{tabular}{cc}
\psfig{file=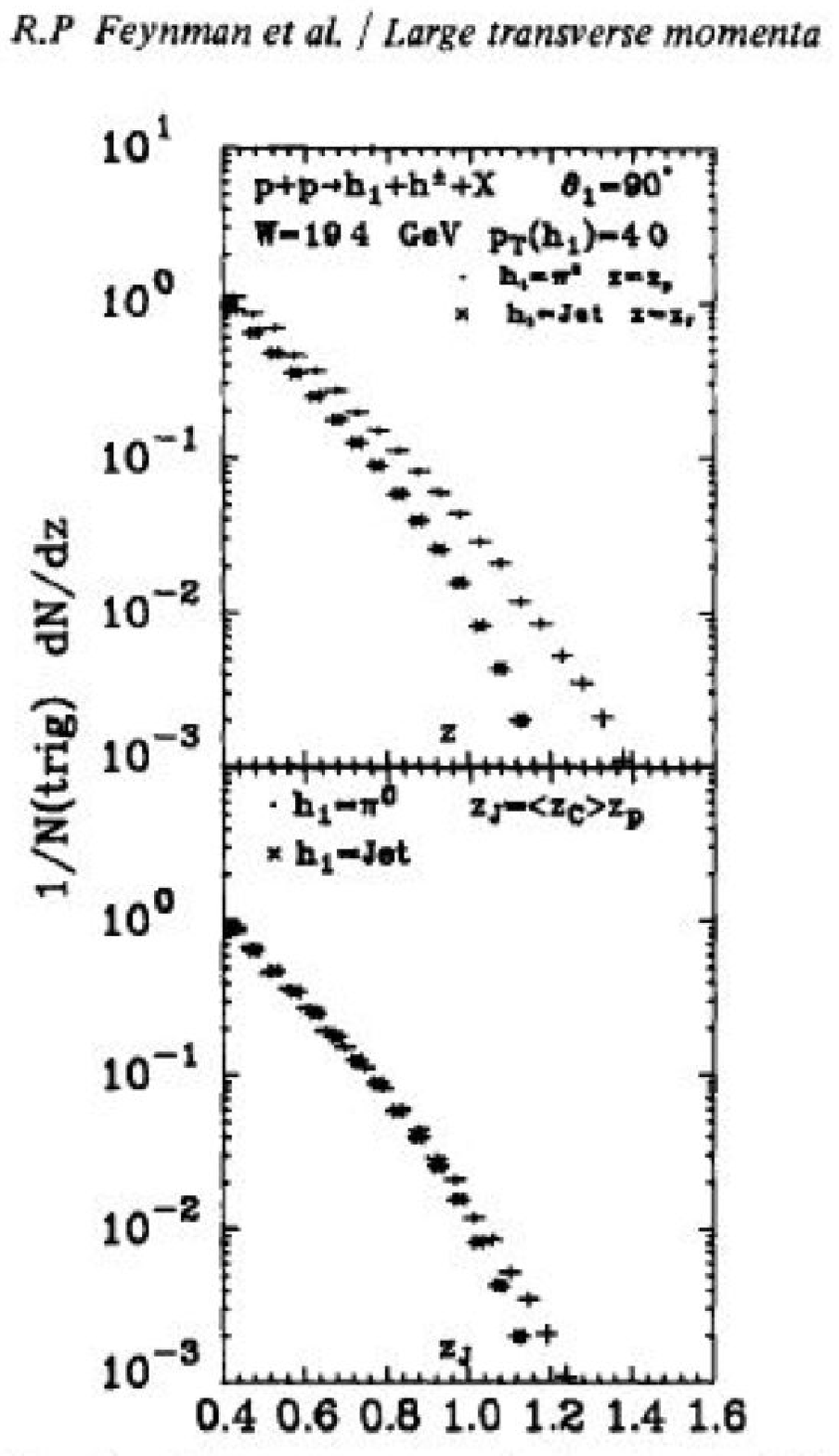,width=0.37\linewidth,height=0.375\linewidth}&
\psfig{file=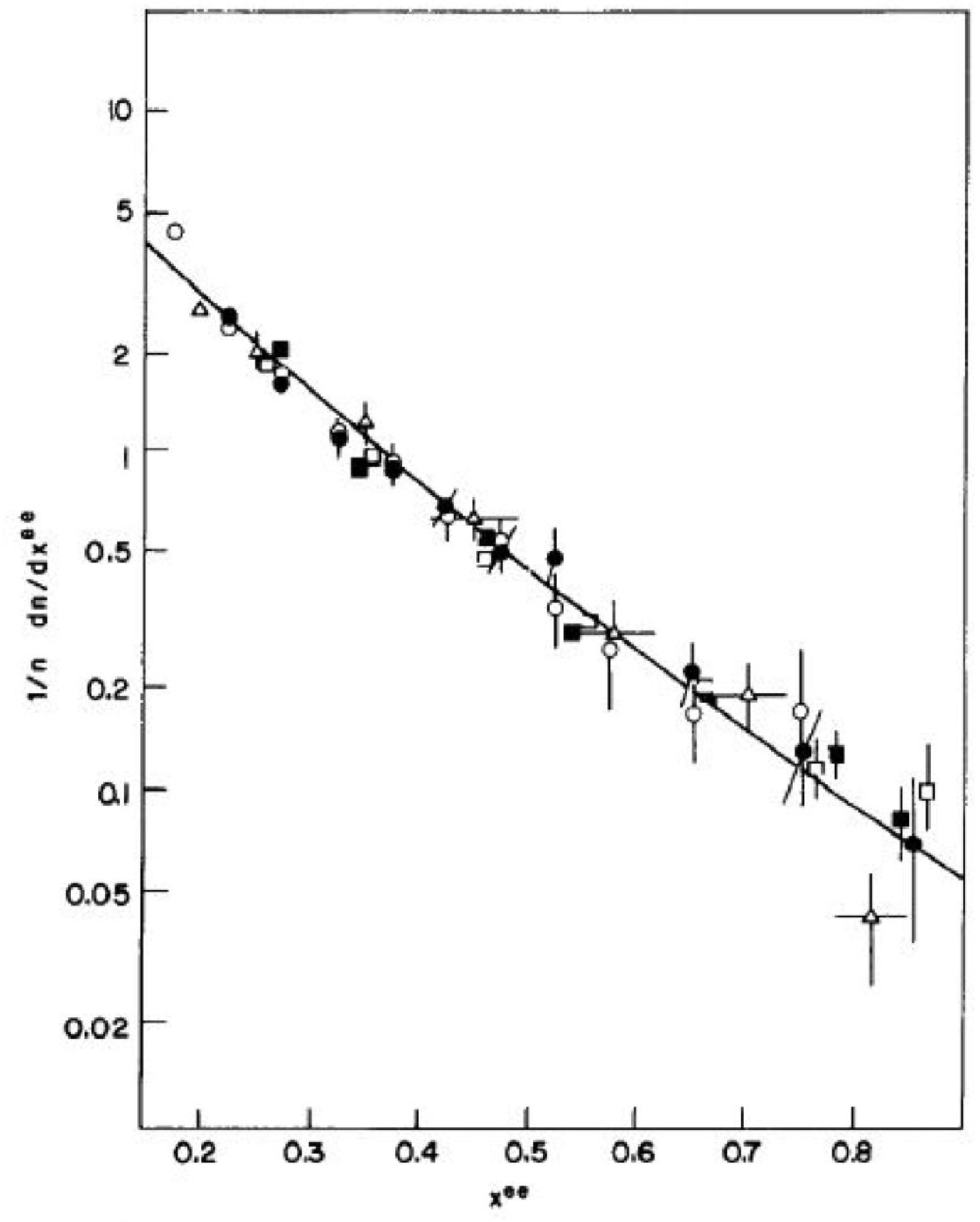,width=0.30\linewidth}
\end{tabular}
\end{center}\vspace*{-0.15in}
\caption[]{a) (left) [top] Comparison~\cite{FFF} of away side charged hadron distribution triggered by a $\pi^0$ or a jet, where $z_{\pi^0}=x_E$ and $z_j=p_{T_a}/\hat{p}_{T_a}$. [bottom] same distributions with $\pi^0$ plotted vs $z'_j=\mean{z_t} x_E$. b) (right) Jet fragmentation functions~\cite{Darriulat} from $\nu$-p, $e^+ e^-$ compared to p-p collisions (Fig.~\ref{fig:mjt-ccorazi}c).    \label{fig:xxx2}}
\end{figure}

\noindent $z_{t}\simeq p_{T_t}/\hat{p}_{T_t}$ is the fragmentation variable of the trigger jet with $\hat{p}_{T_t}$, and $z\simeq p_{T_a}/\hat{p}_{T_a}$ is the fragmentation variable of the away jet. Note that $x_E$ would equal the fragmenation fraction $z$ of the away jet, for $z_{t}\rightarrow 1$, if the trigger and away jets balanced transverse momentum, i.e. if $\hat{x}_h\equiv\hat{p}_{T_a}/\hat{p}_{T_t}=1$.  
It was generally assumed, following the seminal article of Feynman, Field and Fox~\cite{FFF}, that the $p_{T_a}$ distribution of away side hadrons from a single particle trigger [with $p_{T_t}$], corrected for $\mean{z_t}$, would be the same as that from a jet-trigger (Fig.~\ref{fig:xxx2}a) and follow the same fragmentation function as observed in $e^+ e^-$  or DIS~\cite{Darriulat} (Fig.~\ref{fig:xxx2}b). The $x_E$ distributions~\cite{Angelis79} for the data of Fig.~\ref{fig:mjt-ccorazi}b are shown in Fig.~\ref{fig:mjt-ccorazi}c and show the fragmentation behavior expected at the time, $e^{-6\,z}\sim e^{-6\,\langle z_{t}\rangle\,  x_E }$. 

\section{2007-`everything' $\rightarrow$ `almost everything'!}
\begin{figure}[!htb]
\begin{center}
\begin{tabular}{ccc}
\psfig{file=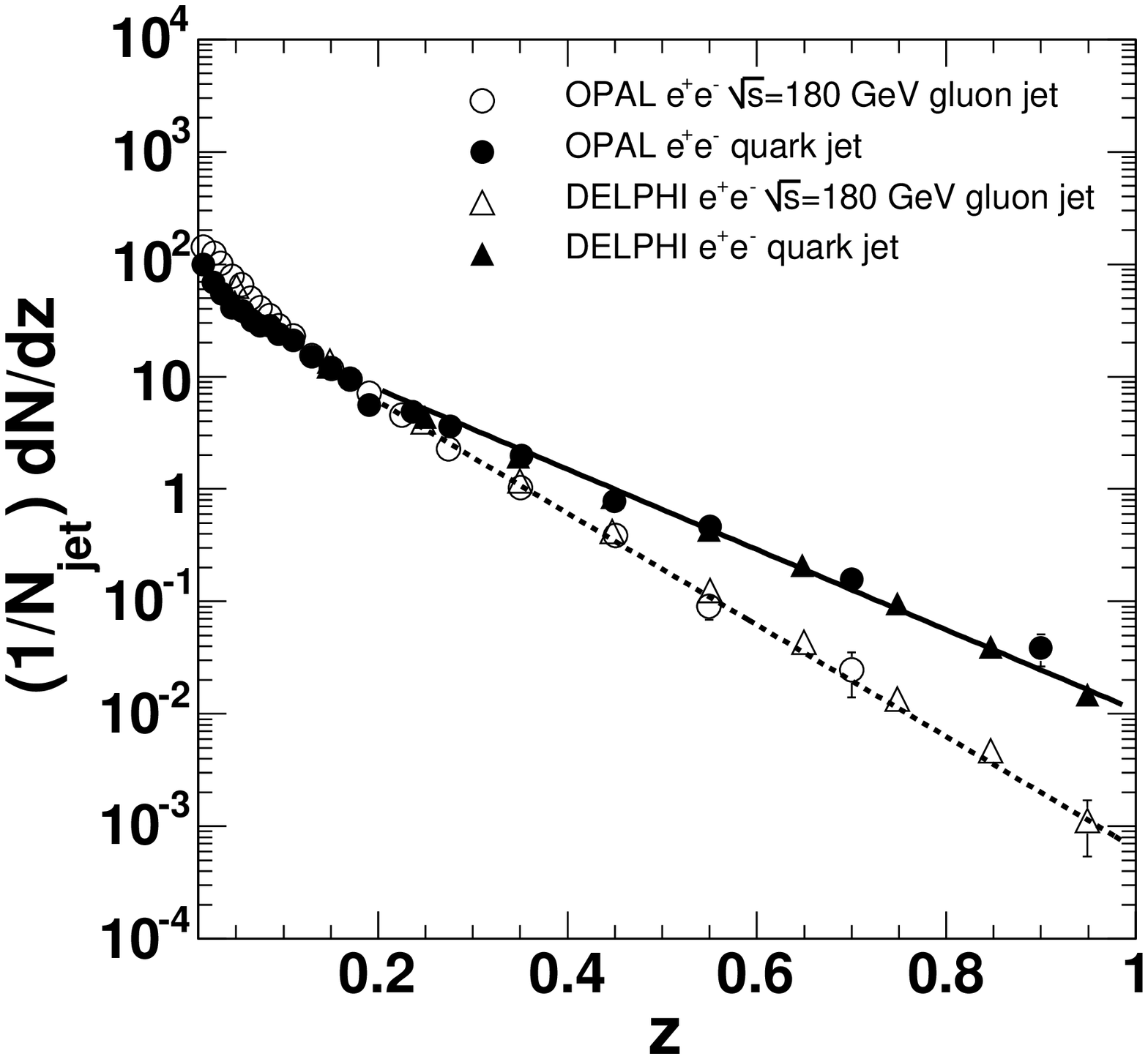,width=0.33\linewidth}&\hspace*{-0.053\linewidth}
\psfig{file=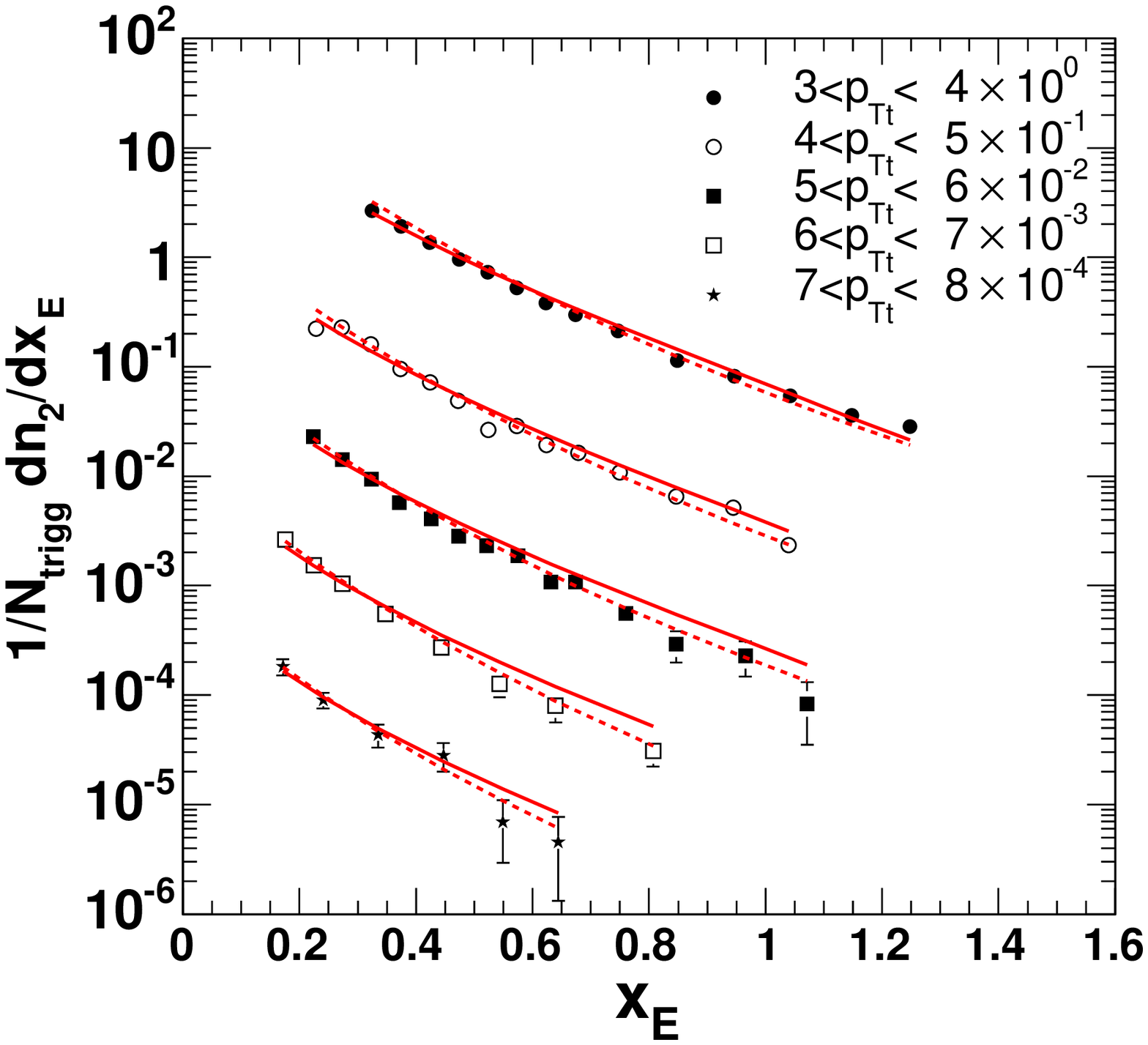,width=0.36\linewidth}&\hspace*{-0.06\linewidth}
\psfig{file=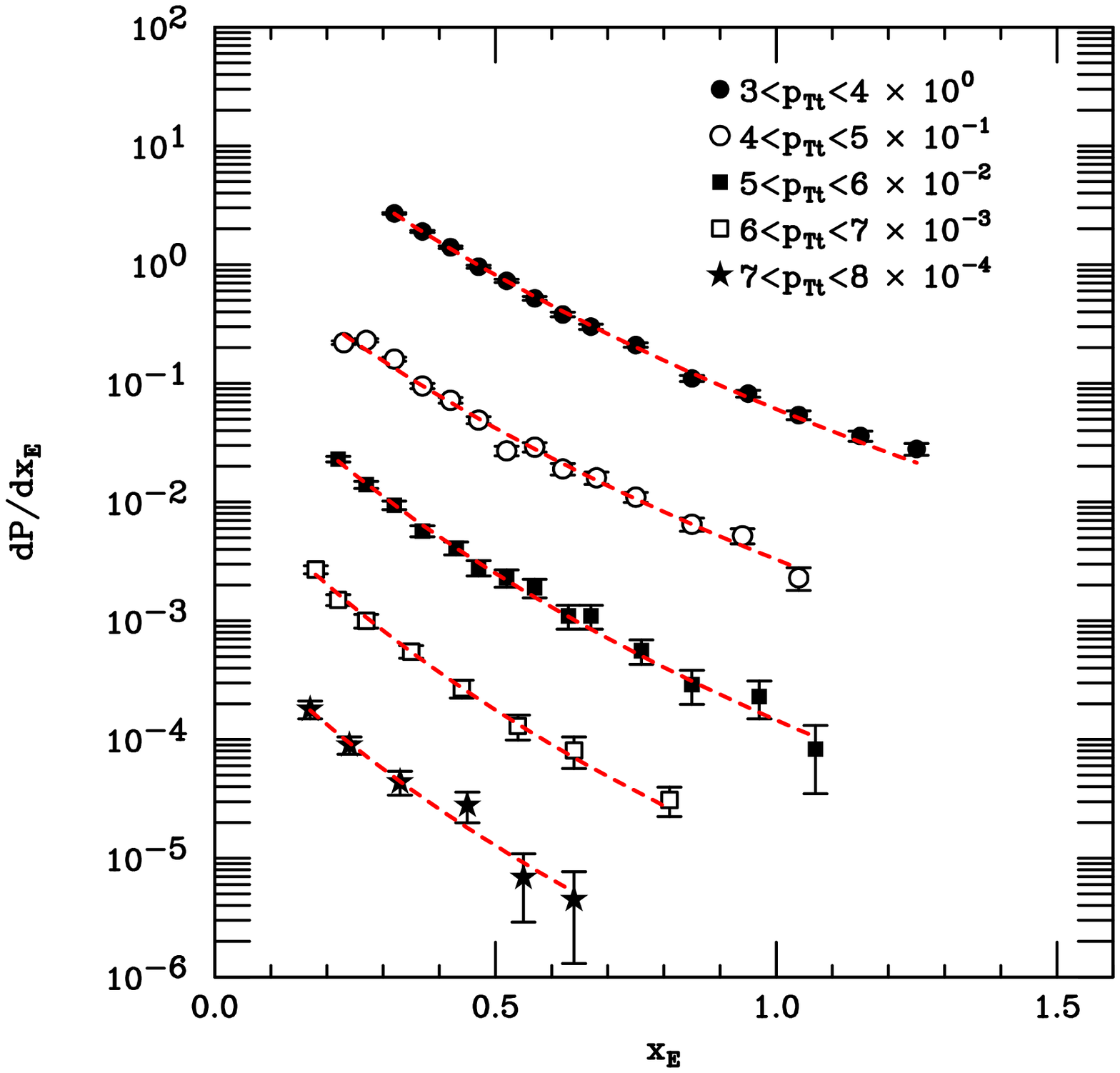,width=0.32\linewidth}
\end{tabular}
\end{center}\vspace*{-0.25in}
\caption[]{a) (left) LEP fragmentation functions. See Ref.~\cite{ppg029} for details; b) (center) $x_E$ distributions~\cite{ppg029} together with calculations using fragmentation functions from LEP; c) (right) data from (b) with fits to Eq.~\ref{eq:condxe2} for $n=8.10$.    \label{fig:xxx3}}\vspace*{-0.12in}
\end{figure}

    PHENIX~\cite{ppg029} attempted to measure the mean net transverse momentum of the di-jet ($\mean{p_{T{\rm pair}}}=\sqrt{2}\mean{k_T}$) in p-p collisions at RHIC, where $k_T$ (see above) represents the out-of-plane activity of the hard-scattering~\cite{FFF,Darriulat}. This requires the knowledge of $\mean{z_t}$ of the trigger $\pi^0$, which PHENIX attempted to calculate using a fragmentation function derived from the measured $x_E$ distributions (Fig.~\ref{fig:xxx3}b).  
It didn't work. Finally, it was found that starting with either the quark $\approx \exp (-8.2 \cdot z)$ or the gluon $\approx \exp (-11.4 \cdot z)$ fragmentation functions from LEP (Fig.~\ref{fig:xxx3}a solid and dotted lines), which are quite different in shape, the results obtained for the $x_E$ distributions (solid and dotted lines on Fig.~\ref{fig:xxx3}b) do not differ significantly! Although nobody had noticed this for nearly 30 years, the reason turned out to be quite simple. The integration over $z_t$ of the trigger jet is actually an integral over the trigger jet $\hat{p}_{T_t}$ for fixed $p_{T_t}$. However since the trigger and away-jets are always roughly equal and opposite in transverse momentum, integrating over $\hat{p}_{T_t}$ simultaneously integrates over $\hat{p}_{T_a}$ and thus also integrates over $z$ of the away-jet. With no assumptions other than a power law for the jet $\hat{p}_{T_t}$ distribution (${{d\sigma_{q} }/{\hat{p}_{T_t} d\hat{p}_{T_t}}}= A \hat{p}_{T_t}^{-n}$) , an exponential fragmentation function ($D^{\pi}_q (z)=B e^{-bz}$), and constant $\hat{x}_h$, for fixed $p_{T_t}$ as a function of $p_{T_a}$, it was possible to derive the $x_E$ distribution in the collinear limit, where $p_{T_a}=x_E p_{T_t}$~\cite{ppg029}: 
	     \begin{equation}
\left.{dP_{\pi} \over dx_E}\right|_{p_{T_t}}\approx {\mean{m}(n-1)}{1\over\hat{x}_h} {1\over
{(1+ {x_E \over{\hat{x}_h}})^{n}}} \, \qquad ,  
\label{eq:condxe2}
\end{equation}
and $\mean{m}$ is the multiplicity of the unbiased away-jet. The shape of the $x_E$ distribution is given by the power $n$ of the partonic and inclusive single particle transverse momentum spectra and does not depend on the exponential slope of the fragmentation function (Fig.~\ref{fig:xxx3}c). Note that Eq.~\ref{eq:condxe2} provides a relationship between the ratio of the away and trigger particle's transverse momenta, $x_{E}\approx p_{T_a}/p_{T_t}$, which is measured, to the ratio of the transverse momenta of the away to the trigger jets, $\hat{x}_h=\hat{p}_{T_a}/\hat{p}_{T_t}$, which can thus be deduced. In p-p collisions the imbalance of the away-jet and the trigger jet ($\hat{x}_h\sim 0.7-0.8$) is caused by $k_T$-smearing~\cite{FFF,ppg029} (Fig.~\ref{fig:xxx3}c).  In A+A collisions, $\hat{x}_h$ is sensitive to the relative energy loss of the trigger and associated jets in the medium, which can be thus measured~\cite{egMJTFI06}. 
\subsection{Suppression of high $p_T$ $\pi^0$ in Au+Au collisions.}
\begin{figure}[!thb]
\begin{center}
\begin{tabular}{cc}
\includegraphics[width=0.38\linewidth]{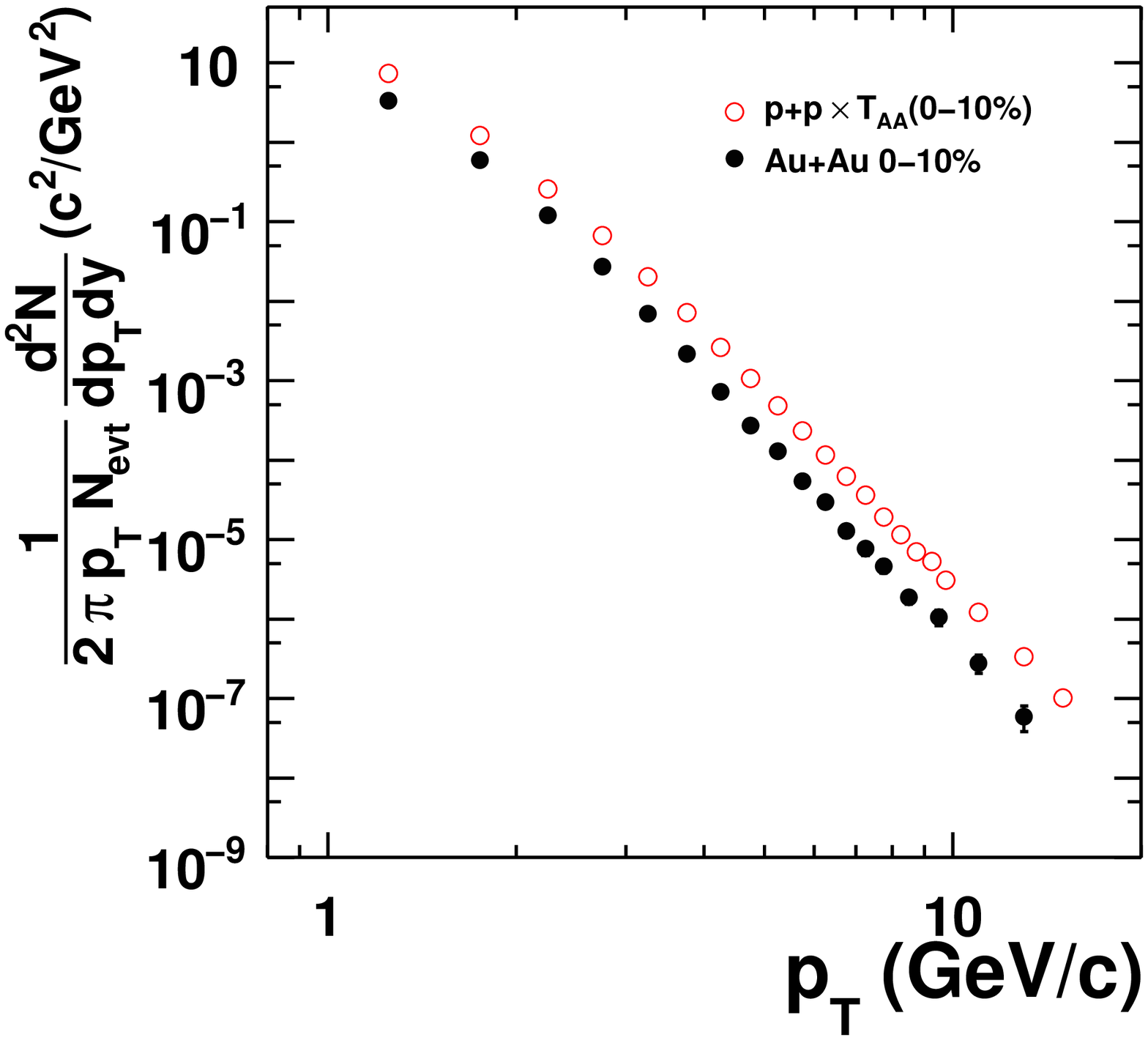}&
\hspace*{-0.02\linewidth}\includegraphics[width=0.60\linewidth]{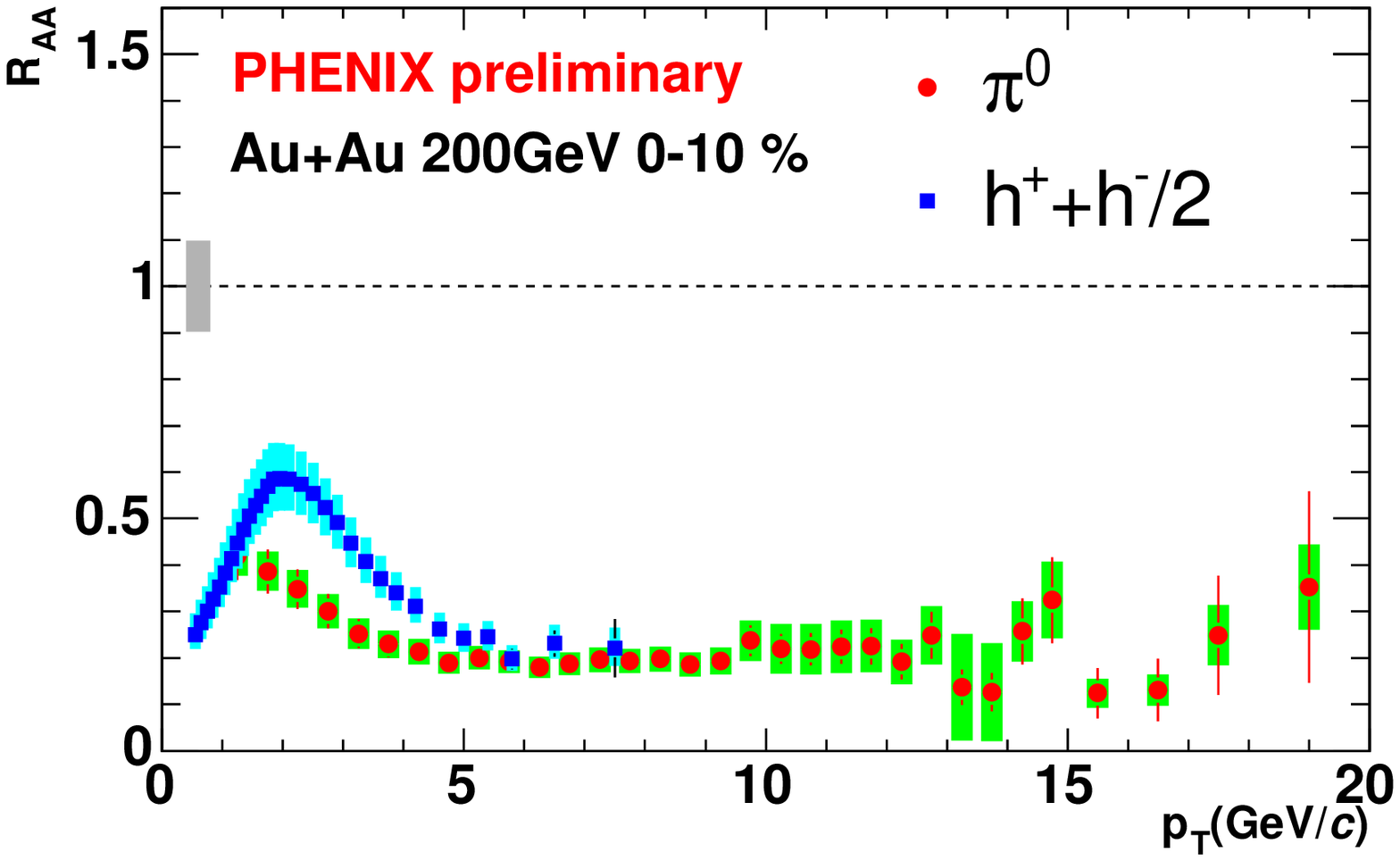}
\end{tabular}
\end{center}\vspace*{-0.25in}
\caption[]{(left)log-log plot~\cite{ppg054} of $\pi^0$ invariant cross section in p-p collisions at $\sqrt{s}=200$ GeV multiplied by $\mean{T_{AA}}$ for Au+Au central collisions (0-10\%) compared to the measured semi-inclusive invariant yield of $\pi^0$. (right) $R_{AA}(p_T)$ for $\pi^0$ and $h^{\pm}$ for Au+Au central (0-10\%) collisions at $\sqrt{s_{NN}}=200$ GeV~\cite{Maya}.  }
\label{fig:RAA}
\end{figure}

        It was discovered at RHIC~\cite{egPXWP} that $\pi^0$ are suppressed by roughly a factor of 5 compared to point-like scaling of hard-scattering in central Au+Au collisions. This is arguably {\em the}  major discovery in Relativistic Heavy Ion Physics. The suppression is attributed to energy-loss of the outgoing hard-scattered color-charged partons due to interactions in the presumably deconfined and thus color-charged medium produced in Au+Au (and Cu+Cu) collisions at RHIC~\cite{BaierQCD98,BSZARNPS}. In Fig.~\ref{fig:RAA}-(left), a log-log plot of the $\pi^0$ invariant cross section in p-p collisions at $\sqrt{s}=200$ GeV multiplied by the point-like scaling factor $\mean{T_{AA}}$ (the overlap integral of the nuclear thickness functions averaged over the centrality class) for Au+Au central collisions (0-10\%) is compared to the measured semi-inclusive invariant yield of $\pi^0$. Both the Au+Au and p-p data show a pure power law, $p_T^{-8.10}$ for $p_T > 3$ GeV/c. The suppression is shown more dramatically in Fig.~\ref{fig:RAA}-(right) where the the data for $\pi^0$ and non-identified charged particles ($h^{\pm}$) are presented as the ratio of the yield per central Au+Au collision  (upper 10\%-ile of observed multiplicity) to the point-like-scaled p-p cross section:
   \begin{equation}
  R_{AA}(p_T)={{d^2N^{\pi}_{AA}/dp_T dy N_{AA}}\over {\langle T_{AA}\rangle d^2\sigma^{\pi}_{pp}/dp_T dy}} \quad . 
  \label{eq:RAA}
  \end{equation}
Since there is no suppression of $\pi^0$ in any measurement in A+A collisions at lower c.m. energies, $\sqrt{s_{NN}}\leq 31$ GeV/c ~\cite{DdE}, the suppression is unique at RHIC energies and occurs at both $\sqrt{s_{NN}}=200$ and 62.4 GeV.  Fig.~\ref{fig:RAA}-(right) also shows that the suppression of non-identified charged hadrons and $\pi^0$ are different for $2\leq p_T \leq 6$ GeV/c. This is due to the fact that baryons are not suppressed for $2\leq p_T\leq 6$ GeV/c~\cite{ppg015}, called the baryon-anomaly, which is still not understood. 
\subsection{Are direct $\gamma$ suppressed?}
   The latest results from RHIC on precision measurements of direct photon production in p-p and Au+Au collisions have solved one mystery but generated another one. The PHENIX measurements of direct photon production in p-p collisions over the range $4\leq p_T\leq 20$ GeV/c at $\sqrt{s}=200$ GeV~\cite{IsobeQM06} ($0.04\leq x_T\leq 0.20$) when combined with the previous data have led to much improved agreement with NLO calculations~\cite{Aurenche06} and confirm that the $\sqrt{s}$ dependence of the
reaction, predominantly $g+q\rightarrow 
   \gamma+q$ can be properly described within the NLO formalism.
\begin{figure}[!thb]
\begin{center}
\begin{tabular}{cc}
\includegraphics[width=0.49\linewidth]{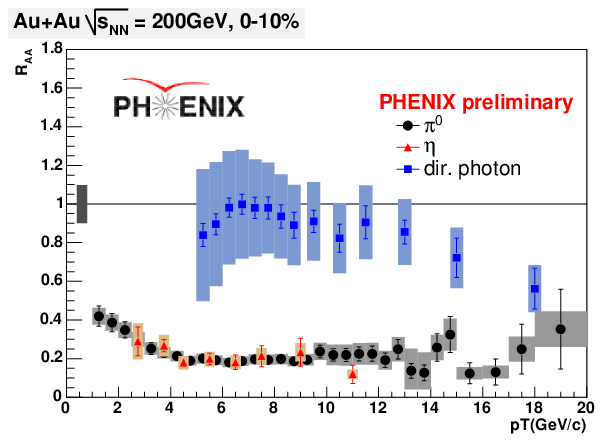}&
\hspace*{-0.02\linewidth}\includegraphics[width=0.49\linewidth]{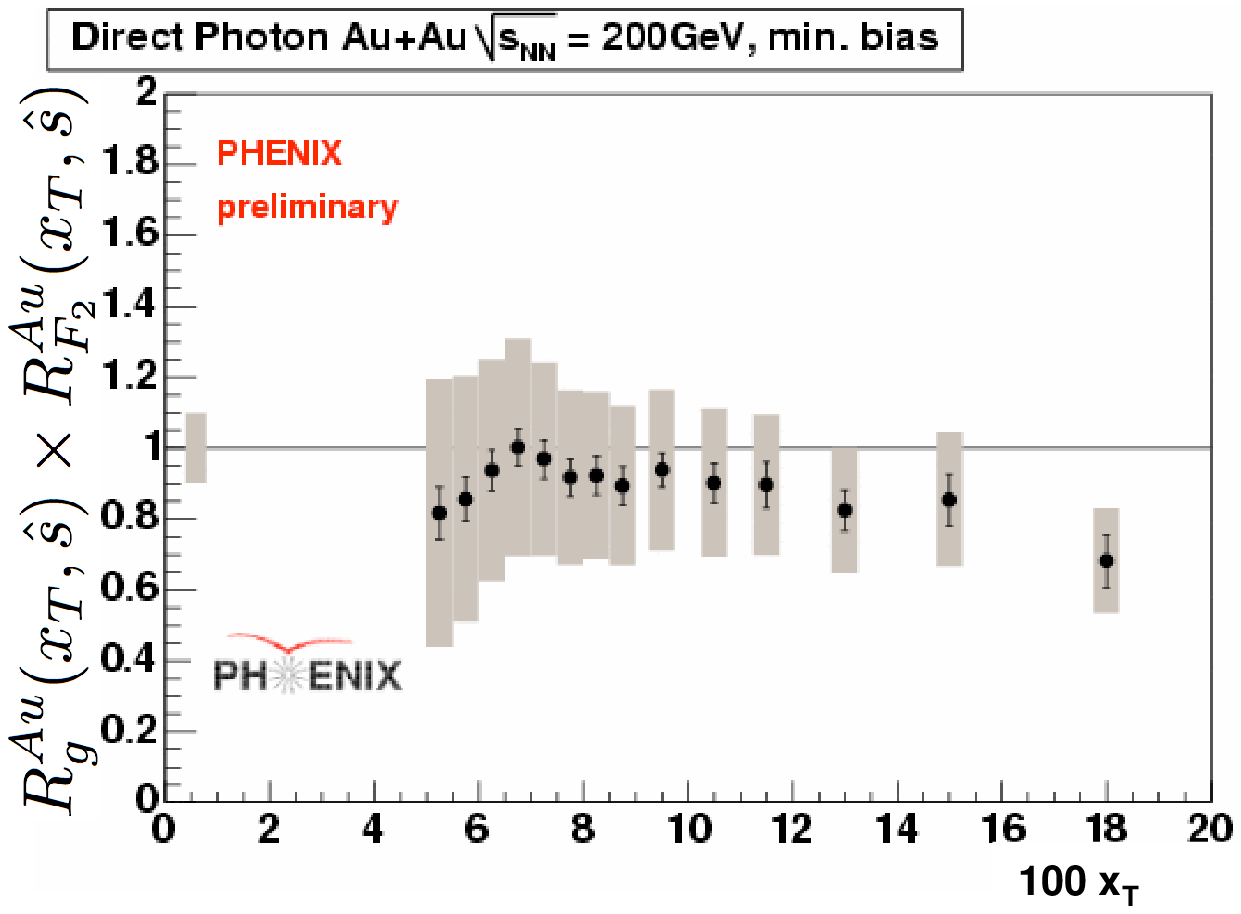}
\end{tabular}
\end{center}\vspace*{-0.25in}
\caption[]{a) (left) $R_{AA}(p_T)$ for direct-$\gamma$, $\pi^0$, $\eta$ for Au+Au central(0--10\%) collisions at $\sqrt{s_{NN}}=200$ GeV~\cite{BaldoQM06}. b) (right) $R^{\gamma}_{AA}(p_T)$ in minimum bias Au+Au collisions at $\sqrt{s_{NN}}=200$ GeV~\cite{IsobeQM06} }
\label{fig:dirG}
\end{figure}
The new mystery~\cite{BaldoQM06} is $R^{\gamma}_{AA}$ shown in Fig.~\ref{fig:dirG}-(left). For $p_T\leq 14$ GeV/c the direct-$\gamma$ are not suppressed, which is expected since they are direct participants in the hard-scatering which do not interact with the medium, while the $\pi^0$ and $\eta$, which are fragments of jets, are suppressed by the same amount. 
This indicates that suppression is a final state effect on outgoing partons. For larger $p_T\sim 20$ GeV/c, the $\pi^0$ suppression is roughly constant at $R_{AA}\simeq 0.2$ while the direct-$\gamma$ appear become suppressed at a level approaching that of $\pi^0$ (with large systematic errors). If this were true, i.e. $R^{\pi^0}_{AA}=R^{\gamma}_{AA}$ for $p_T > 20$ GeV/c, it would indicate that the suppression in this higher $p_T$ range is not a final state effect (i.e. the energy loss becomes negligible compared to $p_T\sim 20$ GeV/c) and must be due to the structure functions. This is easy to understand for direct photons at mid-rapidity in minimum bias Au+Au for which:
\begin{equation}
R^{\gamma}_{AA}(p_T)\approx \left( \frac{F_{\rm 2A}(x_T)}{AF_{2p}(x_T)} \times \frac{g_{A}(x_T)}{Ag_{p}(x_T)}\right) \qquad ,
\end{equation}
where $F_{2p}(x)$ is the structure function from DIS and $g_p(x)$ is the gluon structure function. Unfortunately, there are no structure function measurement as a function of centrality in the 40 years of DIS~\cite{E665}. 
    
    RHIC (and soon LHC) are the only places in the universe where the interactions of color-charged partons in a color-charged medium (produced in A+A collisions) can be studied. This opens a totally new field of fundamental precision QCD studies where surprises may be expected~\cite{Z}. 
 

\begin{thebibliography}{99}
\bibitem{BaierQCD98} R. Baier, {\em QCD, Proc. IV Workshop-1998 (Paris)}  
(World Scientific, Singapore, 1999) pp 272--279   
\bibitem{MJTQCD98}M.~J.~ Tannenbaum,  {\it ibid.}, 
pp 280--285, pp 312--319
\bibitem{Shura2000} A.~Bazilevsky {\it et al.}, {RIKEN Review\ } {\bf 28}, 15 (2000). 
\bibitem{MGyulassy} M.~Gyulassy and M.~Pl\"umer,  \Journal{\PLB}{243}{432}{1990} 
\bibitem{XNWang} X.-N.~Wang  and M.~Gyulassy, \Journal{\PRL}{68}{1480}{1992} 
\bibitem{Darriulat} For example, see P.~Darriulat, \Journal{\ARNPS}{30}{159}{1980}.
\bibitem{Angelis79} A.~L.~S.~Angelis, {\it et al.} (CCOR), \Journal{Physica Scripta\ }{19}{116--123}{1979}. 
\bibitem{FFF} R.~P.~Feynman, R.~D.~Field, and G.~C.~Fox, \Journal{\NPB}{128}{1--65}{1977}.
\bibitem{egPXWP} e.g. see K.~Adcox, {\it et al.} (PHENIX), \Journal{\NPA}{757}{184--283}{2005}.
\bibitem{ppg054} S.~S.~Adler, {\it et al.} (PHENIX), \Journal{\PRC}{76}{034904}{2007}. 
\bibitem{Maya} M.~Shimomura, {\it et al.} (PHENIX), \Journal{\NPA}{774}{457}{2006}.
\bibitem{DdE} e.g. see D.~d'Enterria, \Journal{\PLB}{596}{32}{2004}.
\bibitem{BSZARNPS} See R.~Baier, D.~Schiff and B.~G.~Zakharov, \Journal{\ARNPS}{50}{37}{2000}.
\bibitem{PXxTAuAu} S.~S.~Adler, {\it et al.} (PHENIX), \Journal{\PRC}{69}{034910}{2004}.   
\bibitem{ppg015} S.~S.~Adler, {\it et al.} (PHENIX), \Journal{\PRL}{91}{172301}{2003}.
\bibitem{ppg029} S.~S.~Adler, {\it et al.} (PHENIX), \Journal{\PRD}{74}{072002}{2006}.
\bibitem{egMJTFI06} e.g. see M.~J.~Tannenbaum, PoS (CFRNC2006) 001.
\bibitem{IsobeQM06} T. Isobe, {\it et al.} (PHENIX), \Journal{\JPG}{34}{S1015--S1018}{2007}.
\bibitem{Aurenche06} P.~Aurenche, {\it et al.}, \Journal{\PRD}{73}{094007}{2006}.
\bibitem{BaldoQM06} B. Sahlmueller, {\it et al.} (PHENIX), \Journal{\JPG}{34}{S969--S974}{2007}.
\bibitem{E665} See, however, M.~R.~Adams, et al. (E665), \Journal{\ZPC}{67}{403}{1995}. 
\bibitem{Z} See, for example, A. Zichichi, \Journal{CERN Courier\ }{47}{43--46}{September 2007}, {\tt http://www.ccsem.infn.it/ref/yukawa.html}.
\end{thebibliography}
\end{document}